# WSN Simulators Evaluation: An Approach Focusing on Energy Awareness

Michel Bakni[1], Luis Manuel Moreno Chacón [2], Yudith Cardinale[2], Guillaume Terrasson[1], and Octavian Curea[1]

[1]Univ. Bordeaux, ESTIA Institute of Technology, F-64210 Bidart, France
[2]Universidad Simón Bolívar, Caracas, 1080-A, Venezuela

*ABSTRACT*

*The large number of Wireless Sensor Networks (WSN) simulators available nowadays, differ in their design, goals, and characteristics. Users who have to decide which simulator is the most appropriate for their particular requirements, are today lost, faced with a panoply of disparate and diverse simulators. Hence, it is obvious the need for establishing guidelines that support users in the tasks of selecting a simulator to suit their preferences and needs. In previous works, we pro- posed a generic and novel approach to evaluate networks simulators, considering a methodological process and a set of qualitative and quantitative criteria. In particularly, for WSN simulators, the criteria include relevant aspects for this kind of networks, such as energy consumption modelling and scalability capacity. The aims of this work are: (i) describe deeply the criteria related to WSN aspects; (ii) extend and update the state of the art of WSN simulators elaborated in our previous works to identify the most used and cited in scientific articles; and (iii) demonstrate the suitability of our novel methodological approach by evaluating and comparing the three most cited simulators, specially in terms of energy modelling and scalability capacities. Results show that our proposed approach provides researchers with an evaluation tool that can be used to describe and compare WSN simulators in order to select the most appropriate one for a given scenario.*

## 1. Introduction

In the last two decades, the use of Wireless Sensor Networks (WSN), in monitoring and tracking applications, has gained attention. The flexible network structure and scalable topology provide attractive solutions for both designers and researchers in the area of data networks [1]. However, developing WSN applications involves a conception and designing phase, after which, the test phase takes place. This test phase can be expensive and includes delays in the development of such as WSN applications. Thus, simulators might be used to save cost and time [2]. In this context, research groups have developed different WSN simulators that answer their needs. As a result, many simulators with various capabilities are available. Thus, the selection of a simulator to implement a specific scenario proposes the following questions: How does a user or a researcher select a WSN simulator? On which basis the decision is to be built?. These are unaddressed issues in the WSN research domain.

In previous works, we proposed a methodological approach with a set of criteria aiming at evaluating network simulators. The approach was applied to evaluate and compare two network simulators, namely Packet Tracer and GNS3, and their quantitative and qualitative characteristics were described [3][4]. Later, the methodological approach was extended to include the evaluation of characteristics of WSN simulators, such as scalability and energy consumption awareness [5].

In [5], we also elaborated a study of a state of the art for WSN simulators, in order to identify the most used and cited in scientific articles.

In this paper, we extend our previous works in several aspects: (i) describe in detail the criteria related to WSN aspects; (ii) extend and update the state of the art of WSN simulators elaborated in our previous works to identify the most used and cited in scientific articles; this helps to eliminate authors bias or unawareness of certain simulators; and (iii) demonstrate the suitability of our systematic approach by evaluating and comparing three of the most cited simulators, specially in terms of energy modelling and scalability capacities The application of our systematic approach leads to results that are measurable and comparable, giving a comprehensive overview of simulators features, their advantages, and disadvantages, particularly from the point of view of scalability capacity and energy consumption awareness. In this way, users and researches the most appropriate simulator for a given scenario

## 2. RELATED WORK

The flexibility and validation in model construction offered by network simulation has fostered the research and development of multiple and different simulators. Thus, for selecting an appropriate network simulator for a simulation task or scenario, it is important to have good knowledge of the available simulators, along with their strengths and weaknesses, as well as to ensure that the results generated by the simulators are valid (i.e., how reliable are the models used by the simulators). Particularly, for WSNs it is important to evaluate the scalability and energy consumption awareness of simulators.

To support this selection process, some works have proposed comparative criteria to carry out the evaluation of network simulators. For WSNs, the most recent and cited comparative studies, related to our work, are [6][7][8][9][10][11][12][13][14][15][16][17][18][19][20][21][22][23][24][25]. Most of them propose generic comparative qualitative criteria, not related to energy consumption, such as type of simulator, API, languages supported, platforms supported, licenses, network support type, user interface [7][8][15][16][17][21][22][23][24][25]. Only the works proposed in [17][18] consider quantitative criteria, such as CPU utilization, memory usage, execution time, and scalability. Other studies also consider energy consumption modelling (e.g., wireless propagation, power consumption, battery, topology, antenna, radio propagation, noise, and application modelling) and the challenges that face their implementations [6][9][10][12][13][14]. Few of such works are dedicated to evaluate WSN simulators in function of the energy consumption of each component of the WSN nodes and how they model the energy consumption of each component [11][20].

All these works, mainly evaluate WSN simulators based on a set of qualitative criteria, related and not related to scalability and energy consumption, but they do not establish any methodological process to perform the evaluation. Besides considering these generic qualitative criteria, we also include quantitative criteria, to evaluate any type of network simulators, as well as specific criteria to evaluate WSN simulators, such as scalability and energy consumption awareness. Although, we propose a methodological approach to make such evaluation in a systematic and formal way.

As far as we know, the only work that timidly proposes a methodological process is presented in [19]. However, the proposed guidelines and steps are focused on performing the network simulation, by following these steps: (i) evaluate the simulator based on a set of generic criteria (e.g., general features, visual support, flexibility, user support); (ii) select benchmarks to evaluate the simulated scenarios (e.g., network design, network protocols); (iii) conduct the simulation process; (iv) evaluate and analyse results. This methodological approach is focused on how to perform the simulation process; while our systematic approach, besides of considering such aspect, is aimed to evaluate and compare several network simulators to select the most appropriated for a target simulation scenario.

Thus, up to what we know, our proposed methodological approach is the first study that considers qualitative and quantitative criteria, as well as a systematic evaluation process, intended to be generic, flexible, and suitable to support the selection of the most appropriate network simulator, according to the user preferences and requirements.

## 3. WSN SIMULATORS: STATE-OF-THE-ART

In order to extract the most cited WSN simulators, we present a study of WSN simulators that are used in current research. We first describe the systematic process used to follow such study and propose a categorization of scientific articles intended to describe, present, or evaluate WSN simulators. Finally, we statistically analyse the articles on each category.

### 3.1. Systematic Review

In order to find, select, and analyse the most popular and recent WSN simulators, we have followed a systematic review consisting of three main steps: (i) search of works dealing with WSN simulators; (ii) selection of relevant articles; and (iii) statistically find simulators cited in the set of the selected papers.

For the first step, the search was done on the Google Scholar search engine, which provides links to scientific repositories such as IEEE Xplore, ACM, and Springer. The search was based on tags that included the keywords WSN and simulator, combined with tags related to the focus of the papers, such as Survey, Review, Comparison, Evaluation. We obtained more than 60 scientific articles.

In the second step, we select the most relevant articles related to WSN simulators evaluation, proposal, and comparison. From the more than 60 scientific papers obtained in the first step, some of them do not focus on simulators, but on designing and evaluating WSNs. We select works since 2010 and some older ones that have been widely cited. The final result was 49 relevant papers, categorized according to their main focus: (i) comparison papers, that evaluate and compare simulators; (ii) survey papers, in which authors present a general review of WSN simulators; (iii) simulator specific papers, which introduce the design or features of a particular WSN simulator; and (iv) trend papers, which explain the definitions and trends of how researchers evaluate WSN simulators.

In the third step, we analyse the selected papers and present statistics of referenced WSN simulators on each category.

### 3.2. Categories of Scientific Articles

The selected papers were classified in four groups:

**Comparison papers,** which include comparative studies of WSN simulators, based on self-defined criteria that evaluate the differences among simulators. In [8], authors make a review of some of the open source network simulators (i.e., NS2, NS3, OMNeT++, and JSIM), comparing them according to languages supported, platforms supported, licenses, network support type, user interface, and API. In [17], authors compare NS2, NS3, OM- NeT++, and GloMoSiM. A unified scenario is applied by simulating a MANET routing protocol, in order to measure memory usage, computational time, and scalability, from which NS3 demonstrates the best performance. Similarly, in [7][12][16][18][19][21][26][27], some of popular WSN simulators (NS2, NS3, TOSSIM, OMNeT++, JSIM, Castalia, Qual- Net, EmStar, ATEMU, Avrora, SENS, COOJA, etc.) are described and compared based on the their general characteristics, their merits, and their limitations. The studies pre- sented in [6][15], evaluate more than 20 simulators. In [9], authors make a survey of available tools to evaluate WSN applications. They identify a set of models that are necessary to have in a WSN simulator: wireless propagation model, fine-grained energy expenditure model, non-linear battery model, and application model. In [10], authors compare Castalia, TOSSIM, and NS3 based on the sustainability to test dynamic network reconfiguration protocols. One of the topics that they evaluate is the energy consumption model of the simulators. They identify that the ability to model the RF states of the sensors is important to model the energy of sensors. In [11], authors compare NS2, NS3, TOSSIM, and OMNeT++, focusing on the modelling of the energy consumption. They describe the energy consumption of each component of the WSN nodes and show how the studied simulators model the energy consumption of each component. In [13], researchers evaluate four WSN simulators: NS2, Castalia, TOSSIM, and COOJA. The evaluation is made by following criteria that they define in the paper. One of the aspects they used to make the evaluation is the energy consumption model and the ability to model non-linear batteries. They execute a series of real experiments and calibrate the radio propagation model and the energy consumption model. In [14], authors review and compare the fol- lowing simulators: NS2, OMNeT++, Prowler, OPNET, and TOSSIM. They highlight the features of each simulator in MAC and routing support, energy modelling, and im- plemented Radio Frequencies (RF) models. In [20], authors compare Castalia, MiXiM, TOSSIM, and WSNet, based on topology, antenna, radio propagation, noise, RF, medium access control, and energy consumption modelling.

**Survey papers**, that describe WSN simulators in a general way, but there is no comparison among them. More than ten simulators are described in [22][25][28][29], in terms of type of simulator, API, languages supported, platforms supported, licenses, network support type, user interface. In other work, reusability, availability, performance, scal- ability, support for scripting languages, and GUI are the aspects considered to describe about 15 simulators [30]. Besides these type criteria, testbeds and hardware platforms are also considered in [31], in order to assessing different parameters required by WSN applications. In [32][33], WSN simulators are described and classified according to their type: simulators (based on models) or emulators. The work presented in [34], describe thirty five simulators considering simulation models, emulation, and testbeds. More than thirty simulators are described in [35], according to a classification presented by the au- thors, based on the target function of simulators: emulators, topology control simulators, environment and wireless medium simulators, network and application level simulators, cross-level simulators, NS2 based simulators, OMNeT++ based simulators, and Ptolemy II based simulators). In [36], a review of network modelling and simulation tools is presented, including WSN simulators, such as NS2, OPNET, and GloMoSim. Authors in [2] present a review of several WSN simulation tools. They mostly focus on their suitability for large-scale WSNs.

**Simulator-specific papers**, which focus on describing properties and characteristics of new WSN simulators. In [37], WebShawn, an WSN simulator is presented. It does not require a specific platform because is an online simulator. In [38], authors developed an energy consumption model and presented eMnSiM, an energy-aware simulator for WSN. First, they studied a number of existing energy models and WSN simulators. Then, the simulator was developed, evaluated, and validated as the obtained results were compared to results obtained from existing WSN simulators. According to the authors, eMnSiM proved to be more efficient in terms of execution time and packet delivery ratio as compared to NS2, OMNeT++, and Sensor Simulator. NS2 is the simulator used in [39] to analyse S- MAC and leach in WSNs. An architecture-driven modelling platform, called A4WSN, for analysing and developing WSNs is presented in [40]. In [41], the support for heterogeneous networks in IDEA1, is presented. Solar Castalia is described in [42], a simulator focused on modelling solar energy sources for WSN.

**Trends papers** focused on studying proposed approaches to evaluate WSN simulators and research trends. In [1], a general description of WSNs is provided. Additionally, authors address the issues and challenges facing the proper design and implementation of WSNs. In [43], authors compile a large set of papers of wireless communication-related conferences and review the statistics about the tools (i.e., testbeds and simulators) the researchers use to evaluate their experiments. Additionally, they address the issues and challenges facing the proper use of WSN simulators. They assert that simulators do not reproduce actual environmental conditions of deployed systems, thus experimental testbeds can be developed to replace simulators. In [44], authors discuss topics to consider when addressing IoT issues. They present the research trends on IoT simulators in the last years. To achieve that, authors describe existing tools that are used by researchers to prove and evaluate their findings on IoT research. They claim that more work is needed to conduct large-scale, robust and effective IoT simulation, and prototype evaluations.

### 3.3. Statistical Analysis of Selected Papers

In total, in the selected papers there are 403 citations, distributed among more than 100 WSN simulators. According to the number of citations, simulators are categorized into three groups: (i) Group 1, composed by simulators with more than 16 citations; this group includes 4 simulators; (ii) Group 2, involves all simulators with 10 to 15 citations citations; it contains 11 simulators; and (iii) Group 3, covers all simulators that are cited less than 10 times; it contains 94 simulators. Figure 1 presents the number of citations and the number of simulators of each group.

The total sum of citations for simulators of the Group 1 is 101, which represent 25.06% of the citations distributed in 4 simulators. Group 2 has in total 140 citations, which means the 34.74% of the citations. Group 3 has 162 citations, which represents a 40.19% of the total of citations. Figure 1 shows that Group 3 and Group 2 contain more citations than Group 1. In both cases, those citations are distributed in a larger number of simulators.

Figure 2 shows the number of citations of the simulators of Group 1, in which the most cited simulators are NS2, TOSSIM, and OMNeT++. NS2 is presented in 30 papers, TOSSIM is presented in 27 papers, and OMNeT++ is presented in 25 papers.

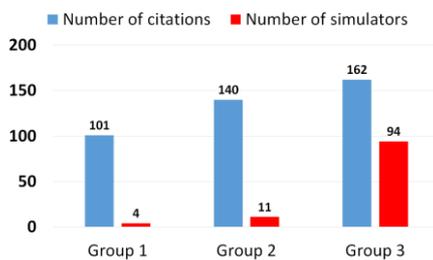
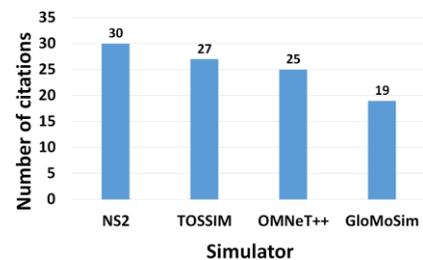

Figure 1: Citations of WSN simulators     Figure 2: Citations in Group 1

This study can help to identify the most used WSN simulators, but the number of citation is not enough to provide comparison-based view. Therefore, a more robust approach to compare and evaluate WSN simulators is needed. In next sections, our methodological approach is presented.

# 4. METHODOLOGICAL APPROACH TO EVALUATE WSN SIMULATORS

In previous works [3][4], we proposed an approach to evaluate data network simulators, considering a methodological process and a set of qualitative and quantitative criteria.

Afterward, we extended our methodological approach to consider qualitative criteria to analyze the simulator scalability and the support of simulators on evaluating traces of energy consumption, sensor nodes mobility, and wireless medium modelling [5]. These characteristics are present in WSN and are less important for general networks. In this work, we continue improving our methodological approach to better considering features that characterize WSN simulators. The methodological process consists on the following steps [3][4][5]:

**Step 1. Establish a set of criteria**. The evaluation of the simulator requires clear and accurate criteria to assess the different aspects of the simulator.
**Step 2. Establish the experimental setup**. The platform in which simulators are installed to be evaluated should not be neglected.
**Step 3. Evaluate the qualitative criteria of the simulator(s)**. To comply this step, it is recommended to revise the available documentation of simulator(s) and elaborate a table highlighting their characteristics.
**Step 4. Design a test scenario to evaluate the measurable criteria.** In a data network and WSN, a scenario is defined by parameters that characterize a specific use case.
**Step 5. Evaluate the measurable criteria of the simulator(s) by executing the designed scenarios**. In order to obtain the results, each designed scenario has to be implemented on the simulator(s).
**Step 6. Elaborate a discussion by analysing the results**.

With this six-steps systematic approach, users can evaluate network simulators to select the most appropriated according to their needs and scenarios. For the comparative analysis, we also propose a set of criteria, which complement the Step 1 of this approach. They are described in the next section.

## 4.1. Criteria used in the Methodological Process

In previous work [5], we define a set of criteria to evaluate WSN simulators, it includes the following items: nature of the simulator, type of simulator, license, user interface, sup- ported platform, level of details, modelling capability, mobility modelling, wireless medium modelling, and Energy consumption modelling.

In this work, we update the level of details item to become design philosophy considering the evaluation of the level of aspects that are being simulated. Each level represents a group of parameters that belong to different functions and features of WSNs. Sorted in descending order, they are abstract algorithms, high-level protocols, low-level protocols, and hardware. The lower the level is, the less the assumptions are and the more the constraints are. If the studied simulator shows only interactions among parameters that belong to the same level, it is a single-level simulator. If it shows interactions among parameters that belong to different levels, but the interactions are limited inside each level, it is a multi-level simulator. In the case where interactions among parameters spread across the levels, it is a cross-level simulator [45].

Additionally, we extended the energy consumption modelling item to include also a description of the energy model using UML diagrams. In general, a class diagram can be used to reflect the structure of the model. However, a state-machine diagram can be designed to reflect the model behaviour.

The methodological approach was applied to the most cited WSN simulators (NS2, TOSSIM, and OMNeT++) in [5]. Although, in this work we extended and updated the state of the art concerning WSN simulators by adding 12 recent studies, these three simulators remain the most cited up to today. The next section presents their comparative evaluation based on our extended approach and focusing on the energy consumption modelling capacity.

## 5. APPLICATION OF THE PROPOSED METHODOLOGICAL PROCESS

To show the suitability of the extended methodological approach, we apply it to compare the most cited WSN simulators, identified in Section 3: NS2, TOSSIM, and OMNeT++.

**Step 1: Establish a set of criteria**. As it is illustrated in Table 1, the set of criteria considered are the ones described in Section 4.2.

**Step 2: Establish the experiment setup**. To evaluate the considered simulators in different systems, they are tested on Linux Ubuntu 16.04 LTS and Microsoft Windows 10 version 10.0.14393. They were installed on the same computer with the following characteristics: Intel(R) Core(TM) i7-7500U CPU @ 2.70GHz with 16 GB of RAM, 915 GB of disk allocated for Linux, while 909 GB is allocated for Windows.

**Step 3: Evaluate the qualitative criteria**. The qualitative criteria of the three simulators are summarized in Table 1. We particularly comment about energy consumption modelling of each simulator.

**NS2 Energy model:** In NS2, the energy model is designed to address the energy consumption in the mobile nodes. The model is a TCL object, it includes variables to store the initial energy of the node and the total energy consumed in transmission, receiving, idle mode, and the sleep mode, as well as the methods intended to deal with these issues. All previously mentioned energy values are stored and treated in Joules.

The model consists of a basic class called Energy Model, and two other subclasses: AdaptiveFidelityEntity and SoftNeighborHandler. The class AdaptiveFidelityEntity manages the energy consumption chronologically, while SoftNeighborHandler is used to control the node relationship with its neighbours. These subclasses are intended to handle events generated in the simulation environment, thus, they inherit the Handler class, which is the base class developed for this purpose. A linked list is used to describe neighbours, where each neighbour forms a node in a series of nodes, with a specific identifier and a timer value to keep the value of the lifetime for the communication channel. Two structures (i.e., struct) were written for that purpose: neighbour_list and neighbour_list_item. It is worth to mention that the information stored for each neighbour does not include its position or coordinates. Figure 3 shows the UML diagram of NS2 energy model.

The model supports four modes of the node: (i) the INROUTE mode, to describe the node while it is moving; (ii) the WAITING mode; (iii) the IDEAL mode; and (iv) the SLEEP mode. The state transition scheme depends on several assumptions, which keep the situation permanently changed. In each state, the energy consumed is calculated based on the power level and the time spent in this state. The power levels can be set as configuration parameters and the time spent in the state is calculated using special functions in the energy model class. The energy consumed is calculated in Joules. Figure 4 shows the machine state diagram for NS2 energy model.

**OMNeT++ Energy model**: The energy model in OMNeT++ consists of three sub- models: storage, generators, and consumption models. These models do not function separately, but they rather operate in parallel. Thus, it is possible to consume stored energy and generate it at the same time. To achieve this, the sub-models relate to each other through a three-level hierarchical structure shown in Figure 5.

Table 1: Comparison of WSN simulators using the proposed criteria

| Criterion | TOSSIM | NS2 | OMNeT++/INET |
|---|---|---|---|
| Nature of the simulator | Emulator | Simulator | Simulator |
| Type of the simulator | discrete-event | discrete-event | discrete-event |
| License | BSD-license | GNU GPLv2 license | Academic Public License. INET models under LGPL or GPL. |
| User Interface | **GUI:** through TinyViz. **Supported languages:** Python, C++ and NesC | **GUI:** through Nam. **Supported languages:** C++ and OTcl | **GUI:** a built-in GUI is available **Supported languages:** C++ and NED |
| Supported platforms | Linux and Windows | Linux, MacOs and FreeBSD | Windows, Linux and Mac OSX |
| Heterogeneity | No | No | Yes |
| Design philosophy | single-level | single-level | single-level |
| Modelling | Available | Available | Available |
| Mobility model | Yes, through MOB-TOSSIM | Yes | Yes |
| Wireless medium model | **Path loss models:** lognormal shadowing **Other models:** noise modelling | **Path loss models:** shadowing, 2-ray ground, free space | **Path loss models:** free-space, log-normal shadowing, rayleight fading, 2-ray ground, rician fading, nakagami fading **Other models:** Background noise, obstacle loss and propagation models |
| Energy model | **Battery model:** No **RF states:** Yes **Limitations:** Cannot model energy harvester units | **Battery model:** Only for Ideal Battery **RF states:** Yes **Limitations:** Cannot model sensing and processing units | **Battery model:** Yes **RF states:** Yes **Limitations:** Cannot model sensing and processing units |
| Supported technology and protocols | TOSSIM simulates entire TinyOS applications, including the network stack that supports TinyOS implementation. | **Application Layer:** DHCP, telnet, FTP, HTTP **Transport Layer:** TCP, UDP, SCTP, XCP, TFRC, RAP, RTPM **Network Layer:** IPv4, IPv6 **Link Layer:** HDLC, GAF, MPLS, LDP, Diffserv, DropTail, RED, RIO, WFQ, SRR, Semantic Packet Queue, REM, CSMA, 802.11b, 802.15.4, Satellite Aloha **Routing Protocols:** RIP, AODV, Click, DSDV, DSR, NixVectorRouting, OLSR | **Application Layer:** HTTP, DHCP, BitTorrent, P2P Video Streaming, Voice **Transport Layer:** TCP, UDP, SCTP, RTP, RTCP. **Network Layer:** ARP, HIP, IGMPv2, IGMPv3, IPv4, IPv6, MCoA, MIPv6 **Link Layer:** 802.11, 802.11p, 802.1e, WiMAX, LDP, LTE, PPP. **Routing Protocols:** AODV, BGP, GPSR, |

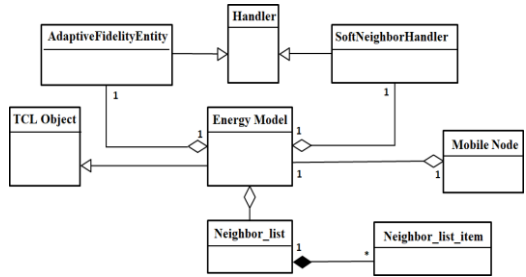 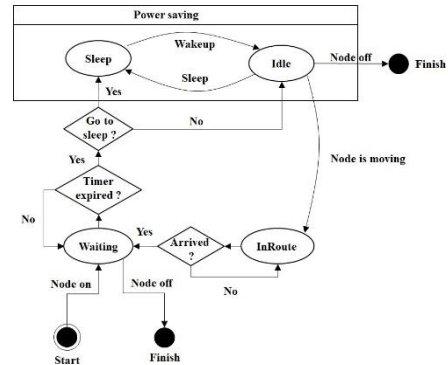

Figure3: UML class diagram–NS2 energy model   Figure4: Machine state diagram–NS2 energy model

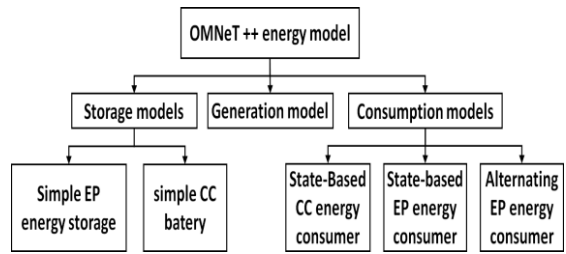

Figure 5: The hierarchical structure of OMNeT++ energy model

At the top level, there are four abstract basic classes: IEnergy Sink, IEnergy Consumer, IEnergy generator, and IEnergy source. Together, they form the basis from which the entire structure is subdivided. In the second level, there is a set of abstract energy classes that connect the abstract classes in the first level with the energy model at the third level. Classes in the third level inherit from the second level the models, the processes, and basic properties associated with energy calculations, whether to make the calculation using energy and power units or using charge and current. In addition, classes of the third level relate to various simulator classes that provide a set of services and basic non-energy functions. Figure 6 shows the UML class diagram of OMNeT++ energy model.

The OMNeT++ energy model supports two forms of calculation, either using power and energy and their units are Watt and Joule respectively, or using charge and current, one of which uses Colomb and Ampere. Although, it is possible to switch from one form to another mathematically, the models in the simulator are built separately. This adds a layer of complexity and causes a repetition of the structure and calculations many times in the code line with only different units used.

**TOSSIM Energy model:** TOSSIM includes the simplest energy model among all the studied models. The model is object-oriented and it does not include a battery model. Instead, it traces the energy consumed by nodes components. The main structure is the struct node energy. Every simulated node has an instance of this struct attached to it. The main struct node energy includes other child data structures: (i) MCUEnergy: this structure is dedicated to the energy consumed by the processing unit; it includes variables to store energy consumed for the microcontroller states, i.e., idle, standby, ex- tended standby, energy-saving, on, and down states. Additionally, the structure includes a variable to trace the energy consumed by the Analog-to-Digital Converter (ADC); (ii) LedEnergy: this is a data structure that traces the energy consumed by the LEDs, where each node can have three LEDs; (iii) RadioEnergy: this structure is used to trace and store energy consumed by the RF module; it follows the state of the circuit. The sending, receiving, and synchronize states are supported; and (iv) MemEnergy: this structure is dedicated to the energy consumed for the memory-related operations; it includes reading and writings. Figure 7 shows the UML class diagram of TOSSIM energy model. The use of the model is simple: every time an activity takes place and consumes energy, the function related to the circuit and the corresponding state is called and the consumed energy is passed to be accumulated in the associated variable.

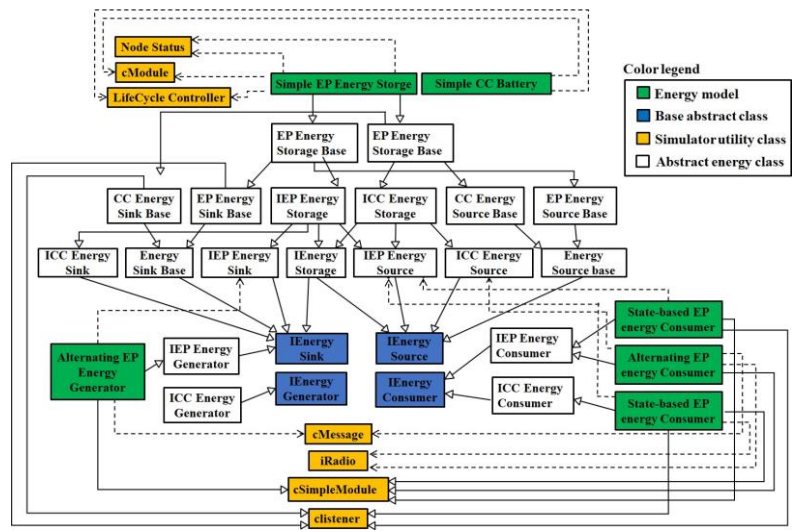

Figure 6: The UML class diagram of OMNeT++ energy model

**Step 4: Design the test scenarios**. Basic scenarios are designed to evaluate the performance of the selected simulators and their energy consumption modelling capacity. The performance is measured in terms of CPU utilization, memory usage, execution time, and scalability. A meshed topology is adopted for the WSN, whose size is increasing exponentially for different tests. The basic component (BC) of the topology consists of four sensor nodes, each one placed in the vertex of a 10x10 meters square. The first test includes only one BC (see Figure 8(a)). The second test is done with two BC, i.e., eight nodes as Figure 8(b) shows. The third one is composed by four BC, with 16 nodes, and so on. In total, eight simulations take place on each system (Linux and Windows), with the number of BCs changing as: 1, 2, 4, 8, 16, 32, 64, and 128 for each simulator.

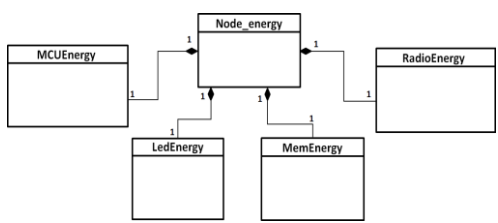

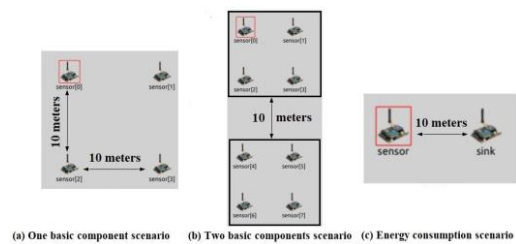

Figure 7: The UML class diagram of
TOSSIM energy model

Figure 8: Simulation scenarios

Each node in the WSN is configured to use IPv4 and ICMPv4. The goal is to create a data message with an echo request to all other nodes in the topology. A node that receives the echo request, replies back the same message. Each simulation lasts 100 seconds. The frequency is 1 Hz, which means that one echo message is sent every second. As a result,

Table 2: Parameters of the energy consumption scenario

| Parameter | 802.11b | 802.15.4 |
|---|---|---|
| Bitrate | 11 Mbps | 250 Kbps |
| MAC | CSMA/CA with RTS/CTS | CSMA/CA with CCA |
| Transmitting power | 750 [mW] | 52 [mW] |
| Receiving power | 220 [mW] | 59 [mW] |
| Sleep power | 0.2 [mW] | 0.06 [mW] |
| Idle power | 0.2 [mW] | 0.06 [mW] |

there are 100 echo request messages sent per simulation.

To evaluate the energy consumption models, another test scenario is proposed. This scenario consists of two nodes, which are 10 meters apart from each other (see Figure 8(c)). One of the nodes is periodically sending an ICMPv4 echo request to the other node. When a node receives the request, it replies back the same message. Both echo request and reply are identical in length and format Therefore, the energy consumption of both nodes will be the same. The communication of nodes is made using two different wireless link protocols: 802.11b and 802.15.4. For each protocol the payload length of the ping message starts at 10 bytes, then, it is gradually being increased by 10 bytes, until the payload size reaches 90 bytes. In total, there are 9 simulation per protocol. Each simulation is repeated three times for different values of frequency of the ping messages: 0.1, 1, and 2 Hz.

For the 802.11b scenarios, the energy consumption parameters were taken from the data sheet of HDG204 RF Module (https://media.digikey.com/pdf/DataSheets/H&DWireless0-PDFs/HDG204DS.pdf), while for the 802.15.4 scenario was used the data sheet of CC2420 RF Module (http://www.ti.com/lit/ds/swrs041c/swrs041c.pdf). Each simulator was con- figured to use the models of the protocols with the values of the standards. The values of the energy consumption for each module is shown in Table 2.

**Step 5: Evaluate the measurable criteria**. NS2 is only evaluated in Linux, since it is the only platform that supports its installation. OMNeT++/INET is installed on both Windows and Linux. The NS2 version used is the 2.35 (https://sourceforge.net/projects/n- snam/), for OMNeT++, it is 5.4.1 (https://github.com/omnetpp/omnetpp/tree/omnetpp- 5.4.1), and for INET, it is 4.1.0-810053f713 (https://github.com/inet-framework/inet/tre- e/v4.1.0). TOSSIM (PowerTOSSIM z) is not installed in none of the systems, since it has a poor level of support for the recent Operating System (OS) versions and it is not possible to install it on the SO used. Thus, PowerTOSSIM z is not evaluated in terms of measurable criteria. We show how the quantitative criteria are measured as follows.

**Performance scenarios**: In these scenarios, the CPU utilization is evaluated for the simulators during 100 seconds of simulation. Results of the evaluation of CPU utilization for different BCs are shown in Figure 9. NS2 tends to consume all available CPU cycles, whatever the number of the BCs is, while OMneT++ consumes the CPU differently in Linux than in Windows. Figure 9 shows that the CPU utilization in Windows is always less than Linux when the same scenario is implemented. In both OS, as the number of BCs increases, the average value of CPU utilization increases as well.

Figure 10 represents the results of memory usage for both simulators on a logarithmic scale as the number of BCs increases. NS2 shows proper memory usage when the BCs are 4 or less. After that, the usage tends to follow an exponential orientation. On both OS, OMNeT++ shows a strictly controlled memory usage as the number of the BCs increases. The memory usage in Windows shows lower values compared to Linux when the same scenario is being implemented.

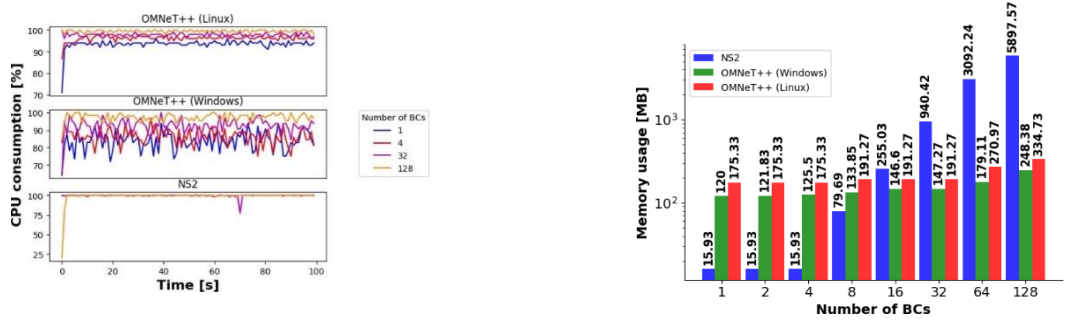

Figure9:CPU utilization of NS2 and OMNeT++   Figure10: Memory usage of NS2 and OMNeT++

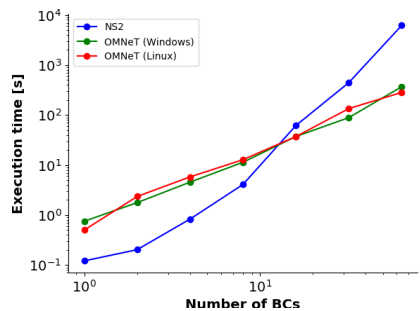

Figure 11: Execution time of NS2 and OMNeT++

To obtain the execution time in OMNeT++, the express-mode is used, since the normal mode was intentionally built to run slowly to allow the user to trace the events that are occurring during the simulation. Figure 11 represents the execution time for the simulators on a logarithmic scale. We note that NS2 has lower execution time for the scenarios with less than 16 BCs, while OMNeT++ has lower execution time for the scenarios that have 16 BCs or more. The execution time of OMNeT++ in Windows and Linux are similar.

Scalability, as the capacity of supporting scenarios with a huge quantity of WSN components, can be deduced from the CPU utilization, memory usage, and total execution time in terms of number of BCs. Results shown on Figures 9, 10, and 11 demonstrate that OMNeT++ scales better than NS2. Even though the CPU utilization of OMNeT++ increases as the number of BCs increases, it is comparable to the CPU utilization of NS2 for the largest scenario (Figure 9), its memory usage increases less than NS2 for larger scenarios (Figure 10), and its total execution time is linear in contrast to the super-linear execution time of NS2 (Figure 11).

**Energy consumption scenarios**: The main objective of the energy consumption scenario is to demonstrate the information that can be obtained from the two simulators. To do so, the same scenarios were implemented on them. In NS2, the entire energy model illustrated in Figure 3 is used. However, for OMNeT++, as shown in Figure 6, the energy model hierarchy includes energy harvester as well, this part of the model was ignored and not used in OMNeT++ simulation to keep the similarity to NS2 simulation, because NS2 does not model energy harvesting.

NS2 has only a command-line interface; thus the output is text displayed on the terminal. Information related to energy consumption is not included. Thus, we developed an animator that was integrated to NS2 as a plug-in, in order to control the simulation time, capture the output, and extract the energy consumption information. OMNeT++ stores information about the simulations in files, that can be exported in multiple formats for later data processing. OMNeT++ shows the same results both on Windows and Linux, regarding the energy consumption evaluation.

International Journal of Wireless & Mobile Networks (IJWMN) Vol. 11, No. 6, December 2019

Therefore, the results of the energy consumption scenario in OMNeT++ are presented only once and without mentioning the OS. The energy model in both simulators trace only the energy consumption of RF module, i.e., the consumption of the node CPU and the sensors are not included.

By comparing the results for the same scenarios obtained from NS2 and OMNeT++, there are differences and similarities. In the 802.11b scenarios, both simulators have the capability to accurately simulate the CSMA/CA mechanism, including parameters of PHY and MAC layers of each frame sent during each phase of the mechanism, such as RTS and CTS frames. Additionally, the data and ACK frames are simulated as well. The implementation uses the standard guideline to define the length of each frame used in the protocol, as well as the preamble length and the PHY header. The time spent sending RTS, CTS, and ACK frames are similar for both simulators as shown in Table 3. But the time spent to send data frames is higher in the OMNeT++ simulator as Table 4 shows.

Table 3: Time spent in 802.11b for control frames

| Frame type | Time in OMNeT++ [$\mu s$] | Time in NS2 [$\mu s$] |
|---|---|---|
| RTS | 207 | 207 |
| CTS | 203 | 202 |
| ACK | 203 | 202 |

Table 4: Time spent in 802.11b for data frames

| Payload [Byte] | Time in OMNeT++ [$\mu s$] | Time in NS2 [$\mu s$] |
|---|---|---|
| 10 | 246 | 239 |
| 30 | 261 | 253 |
| 50 | 275 | 268 |
| 70 | 290 | 282 |
| 90 | 304 | 297 |

By examining one of the repeated interval of the simulation (i.e., the time that includes sending one ping message), the results show that the energy consumption of both sim- ulators are not the same. Figure 12 shows the energy consumption in an interval when the frequency is 1 Hz, for the 802.11b scenario, for both simulators. Each pair of columns represents a payload size; the columns to the right is for results obtained from OMNeT++, while the column to the left is for results obtained from NS2. In general, when the same scenario is implemented, the reported energy consumption in OMNeT++ is slightly higher than NS2 (less than 5%).

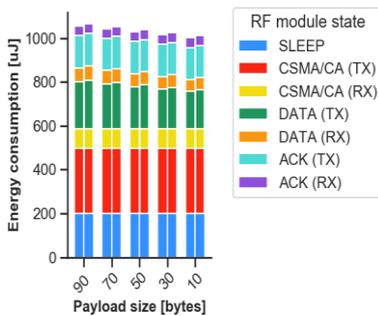
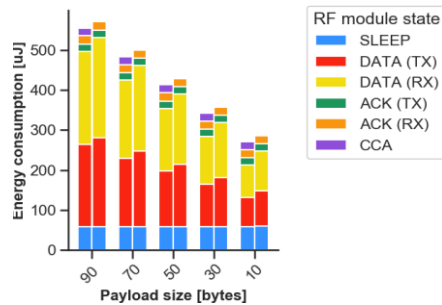

Figure 12: Energy consumption using 802.11b for NS2 and OMNeT++

Figure 13: Energy consumption using 802.15.4 for NS2 and OMNeT++

It is possible to trace the results obtained from the simulators in Table 3 and Table 4, there are different values for the same activities when sending and receiving data frames. These differences between the time spent in sending each frame appears in all the scenarios for 802.11b. It is the cause of the difference in energy consumption between the two simulators.

In the 802.15.4 scenarios, both simulators present different implementations of the proto- col, this, in turn, affects the energy consumption. Figure 13 shows the energy consumption for one interval of 802.15.4, when frequency is 1 Hz.

**Step 6: Elaborate a discussion**. From the methodological process, it is possible to detect advantages and disadvantages of the three analyzed WSN simulators. In the following, the analysis is divided into three sections based on the discussed subject.

**On qualitative criteria:** NS2 is a generic data network simulator that was later adapted to suit WSN, while OMNeT++, was built to support the WSN from the beginning. TOSSIM, is an emulator for TinyOS, which is an OS widely used for embedded systems. The principal drawback of TOSSIM is that it is not compatible with the modern OS. It was not possible to install and run the evaluation scenarios in TOSSIM. Thus, only the qual- itative parameters are available for the comparison with the other simulators. Although, NS2 is only supported on Linux, and despite the fact that it is no longer maintained in favor of NS3, it is one of the most cited simulators in the research domain, meaning that NS3 has not completely replaced NS2. NS3 is still in development and many protocols supported in NS2 have not been yet implemented in NS3. OMNeT++ is supported both on Linux and Windows. The project is still maintained and regularly updated.

**On energy models**: Energy modeling in WSN simulation includes three aspects: the harvesting, storage, and consumption of energy. The studied simulators handle these aspects in different manners.

NS2 does not include energy harvesting. However, storage and consumption are considered and modeled. Energy stored in watts, and each activity taking place in the node consumes a specific amount of energy that is calculated based in different parameters. For example, when sending data packets, bit rate and payload lengths impact energy consumption. The machine state, shown in Figure 4, controls the the sequence of the activities.

OMNeT++ has an object-oriented energy model that links together different models for harvesting, storage, and consumption of energy. As illustrated in Figure 6, the structure of the model allows the creation of different models under each category. There are two possibilities for the units to be used: charge and current or power and energy. The user is to choose a possibility before running the simulation. Then, the transformation is not possible. Both in NS2 and OMNeT++, the consumption model is limited to the RF module, thus sensors processing units activities are not included, although they are responsible for a considerable amount of consumed energy [46].

TOSSIM model is simple and easy to use, but it has some drawbacks. First, the design includes the use of different functions to provide the same service, the only difference is the corresponding circuit, i.e., each circuity has its own energy consumption function, although these functions are written in the same way and provide the same service. In this case, it is recommended to use inheritance and write the function once, then inherit it by other classes. Second, back to the Figure 7, the model includes objects for MCU, LEDs, memory, and RF, but not sensors. The sensors are not involved, thus, their consumption was neglected. Third, and the most important, is the lack of the battery model leads to the need to use a specific amount of energy for circuit activities. This design is not flexible, because parameters such as bit rate, payload, and node positions cannot impact the consumed energy. In other words, the transmission of data packets in different sizes or using different bit rates will end consuming the same amount of energy.

The application of our methodological approach show that none of the studied simulators considers voltage regulators in the model design. Regulators reside between battery and node circuits aiming at providing the stable and level-controlled power transmission. The selection of voltage regulators impacts energy consumption. Thus, it is recommended for the battery model to include parameters that reflect regulator effect. Our criteria do not address the reliability of the energy model, i.e., how close are the obtained results to the real-world tests? It is recommended to extend the criteria to include a set of steps for validating the obtained results with reference scenarios or measurements.

**On performance:** This methodological approach allows studying and evaluating the performance of WSN simulators from three points of view: the CPU consumption, memory usage, and execution time. Additionally, although scalability was included in the proposed criteria, outlines can be drawn for this item by analyzing the results obtained from the previously mentioned items.

**-CPU consumption:** NS2 shows the same behavior regardless of the number of nodes, it uses all available CPU cycles. This leads to a very low execution time when the number of nodes is small. On the other hand, OMNeT++ is more conservative. In Windows, the CPU consumption starts around 70% for scenarios with a small number of nodes. Then, the consumption grows as the number of nodes increases. In Linux, OMNeT++ shows similar behavior except for the starting threshold raises to 90%. In general, from a CPU consumption point of view, OMNeT++ in Windows consumes the CPU the least, when the same scenarios are being executed.

**-Memory usage:** as the number of nodes is increased, the memory usage shows growth for the two simulators. NS2 presents stable usage of memory when the number of nodes is less than 16. Then, the behavior changes and exponential growth takes place. Based on the testing simulation platform we used, it was not possible to simulate scenarios with more than 512 nodes in NS2; this is mainly due to the long execution time obtained which lasts for several days when exceeded the 512 nodes threshold. However, although not presented in this paper, OMNeT++ performs the same except for the threshold that rises to 2048 nodes. Interestingly, memory usage for OMNeT++ is identical when the same scenarios are implemented in two OS that use different architectures. Briefly, in this set of tests, NS2 uses less memory than OMNeT++ when the number of nodes is less than 8, but it uses more memory when nodes' number exceeds this threshold.

**-Execution time:** the execution time for the scenarios showed exceptional growth as a function of the number of nodes. NS2 benefits from its high consumption of the CPU when the number of nodes is low, the scenario executes fast that there are difficulties to obtain the execution time in these scenarios. However, as the number of nodes is increasing, execution time grows in NS2 to exceed that of OMNeT++ in all scenarios when the nodes are more than 16. In NS2, when nodes are more than 512, the simulation is impractical as it lasts for days. On the other hand, simulations in OMNeT++ show close execution times regardless of the OS that hosts the simulator.

**-Scalability:** analyzing the performance criteria, we can highlight scalability. First, memory usage puts strict constraints on the number of nodes used. Information for each simulated node is stored in memory, but simulators vary in the level of details they use. Thus, simulators running the same scenario use different amounts of memory. However, results on memory usage measurements show that when the exponential growth starts, the scalability limit is achieved. Harmoniously with that, results on execution time measurements reflect that there is a threshold for the maximum number of nodes where execution time is practical. Beyond this limit, the simulation can last for days. In general for the proposed scenarios, OMNeT++ is more scalable than NS2. However, dedicated scenarios are needed to be developed in order to address this item in detail.

## 6. DISCUSSION ABOUT THE PROPOSED METHODOLOGICAL APPROACH

Nowadays, with the huge variety of available simulators, it is important to identify which simulator suits the most for a given scenario. The problem of selection always arises, no matter if the simulator is going to be used for academic purposes or industrial development.
From previous proposals [3][4][5], we add extensions aimed to provide the approach with criteria to address the evaluation of WSN simulators, in terms of their scalability and capability of modelling mobility, wireless medium, and energy consumption.

In a WSN simulation environment, energy issues can be addressed using modelling technologies. On the other hand, scalability is a subject governed by the hardware of the simulator host (CPU and memory). Our proposed approach addresses these issues: it proposes guidelines and criteria to evaluate the energy consumption awareness modelling of simulators and to measure their scalability.

Most WSN simulators model the energy consumption of the RF module. Although the RF activities are responsible for the major part of the energy consumption in the node, the consumption of CPU and sensors cannot be neglected. In [46], authors calculate the power consumption average of the sensor unit, the RF module, and the microcontroller for a WSN application. In their specific application the average of power consumed for the RF activities were 62%, the average of power consumed for the sensor and the microcontroller were 14% and 24% respectively; which means that the RF activities can consume more than the sum of the other units. Therefore, it is important for a simulator to model the energy consumption of all units present in the node in order to get an accurate estimation of the energy consumed. Our methodological approach evaluate all these aspects.

The proposed methodological approach is flexible, allowing to integrate another items to cover new aspects needed by users. For instance, it is possible to add criteria to evaluate the simulators capacity of modelling the antenna or the battery behaviour. By following the systematic process steps, the advantages and disadvantages of one or more simulators for a certain application can be identified. Thus, the selection of one of them can be well justified and probed, as well as its suitability for specific user needs and scenarios.

Although the systematic approach provides a comprehensive method to compare WSN simulators, there are still aspects to be covered. For example, the study of energy modelling can be extended to include the support for the battery model. When considering the estimation of the node lifetime, the model that traces the remaining energy is different from the one that traces the consumed energy. The support of parallel processing is another item that can be extended as well. This feature exists in some simulators and has a huge effect on performance. Besides, wireless link protocols have special role in WSN. Thus, it is recommended to separate it from the protocol items and consider additional aspects that concerns to the support of different bit rates and fragmentation.

Finally, WSNs are still in developing and new technologies will be adapted. Thus, new features will be added and WSN simulators have to answer to that. As a result, there is no fixed approach to address WSNs simulators, as they are still constantly changing, the approach that describes and evaluate them will be continuously modified. Our ap- proach faces all these challenges by being extensible, flexible, and generic, and still being a powerful tool to evaluate and compare network simulators.

# 7. CONCLUSIONS AND FUTURE WORK

We have addressed the difficulty of selecting a WSN simulator to fit a given scenario. To achieve that, we extend our previous proposed methodological approach, by integrating new criteria to address WSN evaluation, such as scalability and the modelling of mobility, wireless medium, and energy consumption. In order to demonstrate the efficiency and suitability of our systematic approach, we elaborate the state of the art of WSN simulators, following a systematic review of most cited and recent scientific papers. From this review, we select the three most cited WSN simulators (i.e.., NS2, TOSSIM, and OMNeT++) to evaluate and compare them following our proposed methodological approach. The application of the methodological process proves that it does not only highlight general aspects of the simulators behaviors but it shows their disadvantages as well.

We plan to include other evaluation criteria, such as the capacity of simulators for parallel processing and support of different bit rates and fragmentation. We are also working on proposing an energy consumption model to include the support for the battery behaviour modelling and on the implementation of a recommender system to support the selection.

## ACKNOWLEDGEMENTS

The work presented in this paper has been financially supported in part by the Regional Council of New Aquitaine (as part of the Call for Projects 2016 funds), in the frame of the OUDINI research project.

## AUTHORS

**Michel Bakni** received the B.S. degree in telecommunication and electronics from Tishreen University, Lattakia, in 2013 and the M.S. degree from UTBM, France, in 2017, in mobile and distributed networks. He is currently pursuing the Ph.D. degree with the Doctoral School of the University of Bordeaux (UBx) and at ESTIA. His research interests include Simulation, Wireless Sensor Networks, and Energy consumption optimization. 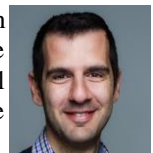

**Luis Manuel Moreno Chacón** is graduated in Telecommunications Engineering at Universidad Simón Bolívar, Venezuela, in 2019. His main areas of research interest are operating systems, distributed systems and embedded systems. 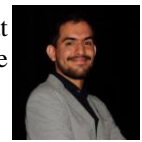

**Yudith Cardinale** is a Full Professor in Computer Science Department at Universidad Simón Bolívar (USB) since 1996. She graduated with honours in Computer Engineering in 1990 at Universidad Centro-Occidental Lisandro Alvarado, Venezuela. She received her M.Sc. Degree and Ph.D. in Computer Science from USB, Venezuela, in 1993 and 2004 respectively. Her research interests include parallel processing, distributed processing, operating systems, high performance on grid and cloud platforms, and web services composition. 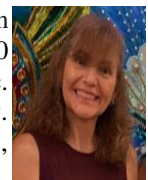

**Guillaume Terrasson** received the M.S. degrees in Microelectronics and the Ph.D degrees in Electronics from the University of Bordeaux 1, France, in 2004 and 2008 respectively. The Ph.D. was obtained in collaboration with the ESTIA and the TIMA laboratory, Grenoble, France. Since 2008, he is a researcher at ESTIA RECHERCHE. 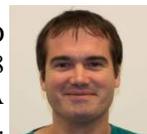

His research interests include critical embedded systems applied to aeronautics as well as wireless sensor networks.

**Octavian Curea** received the Engineer degree in Electrical Engineering at Polytechnic Institute Traian Vuia of Timisoara, Romania, in 1994. He pursued his studies in France by obtaining the M.Sc. degree in 1997, and the Ph.D. degree in 2001, at University of Le Havre, GREAH laboratory. In 2004 he joined ESTIA as associated profesor and he became full profesor in 2017. His research interests include the electrical microgrids from the point of view of energy management, communication networks, power electronics, power quality. 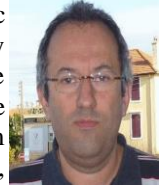